# Automated Space/Time Scaling of Streaming Task Graph

Hossein Omidian
Department of Electrical and Computer
University of British Columbia
Vancouver, Canada
hosseino@ece.ubc.ca

Guy G.F. Lemieux
Department of Electrical and Computer
University of British Columbia
Vancouver, Canada
lemieux@ece.ubc.ca

*Abstract*—**In this paper, we describe a high-level synthesis (HLS) tool that automatically allows area/throughput trade-offs for implementing streaming task graphs (STG). Our tool targets a massively parallel processor array (MPPA) architecture, very similar to the Ambric MPPA chip architecture, which is to be implemented as an FPGA overlay. Similar to Ambric tools, our HLS tool accepts a STG as input written in a subset of Java and a structural language in the style of a Kahn Processing Network (KPN). Unlike the Ambric tools, our HLS tool analyzes the parallelism internal to each Java "node" and evaluates the throughput and area of several possible implementations. It then analyzes the full graph for bottlenecks or excess compute capacity, selects an implementation for each node, and even considers replicating or splitting nodes while either minimizing area (for a fixed throughput target), or maximizing throughput (for a fixed area target). In addition to traditional node selection and replication methods used in prior work, we have uniquely implemented node combining and splitting to find a better area/throughput trade-off. We present two optimization approaches, a formal ILP formulation and a heuristic solution. Results show that the heuristic is more flexible and can find design points not available to the ILP, thereby achieving superior results.**

## I. Introduction

Since clock frequency scaling has essentially stopped due to power issues, the research community has focused on delivering increased levels of parallelism to improve both performance and performance per Watt [1]. At one end, coarse-grained parallelism is achieved with multi-core processors, usually through a high-level language. At the other end, fine-grained parallelism is achieved in FPGAs and ASICs by designing hardware-level solutions. Both of these extremes can be challenging to program. In software, it can be challenging to expose sufficient parallelism in C, and difficult to describe some types of computation in CUDA or OpenCL. In hardware, RTL languages such as VHDL and Verilog achieve very good results, but it is tedious to describe everything at such a low level on a cycle-by-cycle basis. High-level synthesis (HLS) tools that convert C to RTL are very compelling, but impose their own challenging constraints writing C and what can be parallelized [2,3].

In this work, we wish to explore the design space in between traditional multi-core CPUs and low-level FPGA/ASIC solutions. In particular, we will investigate whether an array of ALUs or very lightweight processors, described best as a massively parallel processor array or MPPA, can achieve sufficient levels of performance, and make design entry sufficiently easy, to make them an interesting alternative to more traditional design methods.

To explore the MPPA as an alternative target, we need three things: (a) a tool flow that can compile algorithms into the target, (b) a detailed architecture description or implementation, and (c) a set of benchmarks to compile into the framework. We decided to work on the tool flow first, with the expectation the tools would help us explore a range of architectures. However, we need an initial programming model and an initial target architecture to focus the tools. The programming model should support the strengths of FPGAs, especially pipelined parallelism. Although OpenCL and CUDA allow a programmer to specify thousands of threads, where one thread is essentially "pipelined" behind another one in a streaming fashion, we find that performance can break down when these threads must share information. Thus, we decided to start with the explicit streaming model and architecture that was defined by Ambric [4,5].

In the Ambric model, a Java object is created for each thread, becoming nodes that communicate together through explicitly defined blocking FIFO communication channels. The node can be a primitive node such as an operation or a composite node such as a thread with more than one primitive node. The objects and channels are placed and routed onto an array of 336 processors with a mesh NoC. Each object contains local state and a processing thread. Processing in an object can be variable latency, but computation between objects is synchronized through the blocking FIFOs. Objects may be replicated, thus facilitating some re-use of a program, but all instances are explicitly allocated and defined by the programmer at compile-time. This is very similar to a Kahn Processing Network (KPN) [6], except that in a KPN the FIFOs are assumed to be infinitely deep. The resulting process network exhibits deterministic behaviour that does not depend on the various computation or communication delays.

One of the drawbacks of the Ambric framework is the need for explicit allocation of all objects and channels. The number of objects, and the computational delays within each object, define amount of parallelism and the throughput of the application. Thus, scaling a program to a larger or smaller processor array requires manually re-programming all objects and channels. For the Ambric commercial solution consisting of a single device, this is an acceptable trade-off. However, for a research platform, we must investigate a variety of array sizes, as well as simpler or more complex processors, which requires automatically transforming a streaming application to use more or less space, thereby increasing or decreasing throughput.

In this paper, we describe the beginning of a high-level synthesis tool that can perform such automated space/time tradeoffs. The user describes an initial program in Ambric-style Java, and then defines either a throughput target, or an area budget. The HLS tool analyzes the processing rate of each object (or thread), and propagates these throughputs across the entire computational graph (defined by the communication channels). It also analyzes each thread to determine the degree of internal parallelism. Using this information, it transforms the compute graph to meet the area or throughput target. There are a variety of transformations such as replicating objects (requiring a split/join on the data), subdividing objects into a deeper pipeline (increasing throughput), and merging objects together (decreasing area). At all times, a whole-program approach is taken to optimization, so portions of a program that are not performance-critical will be merged to use less area, and more area will be allocated to performance-critical regions. This alleviates some effort from the programmer, and creates a scalable/retargetable implementation.

Our tool uses two internal optimization approaches. The first, based upon integer linear programming (ILP), is similar to previous work on task graph optimizations by Cong et al [7]. The second, based upon a heuristic approach, is our own novel contribution. Although the ILP approach works well, maintaining the ILP optimization model within the tool prohibits the use of certain optimizations. Instead, our heuristic approach is able to perform object coalescing, which cannot be done as an ILP formulation. This leads to considerable area savings versus the ILP approach.

## II. KPN-BASED HLS FOR MPPA OVERLAY

Figure 1 illustrates the detailed flow of proposed tool. It compiles a program described in Ambric-style Java (compute) and aStruct (communication), and makes a STG complete with composite nodes communicating with each other through channels (edges of the graph). It can handle STGs without any feedback for now but we will address handling STGs with feedback as our future work. Our tool uses Intra-Node Optimizer and Inter-Node Optimizer in order to find different implementations for each composite node. It uses Trade-off Finder to find a good trade off between throughput and area. Later, we will implement MPPA Overlay Generator to generate a suitable architecture instance based on the found trade-off and different types of PE in an Architecture Library. A Back-end module will also be implemented to generate MPPA instructions and map the application to the overlay.

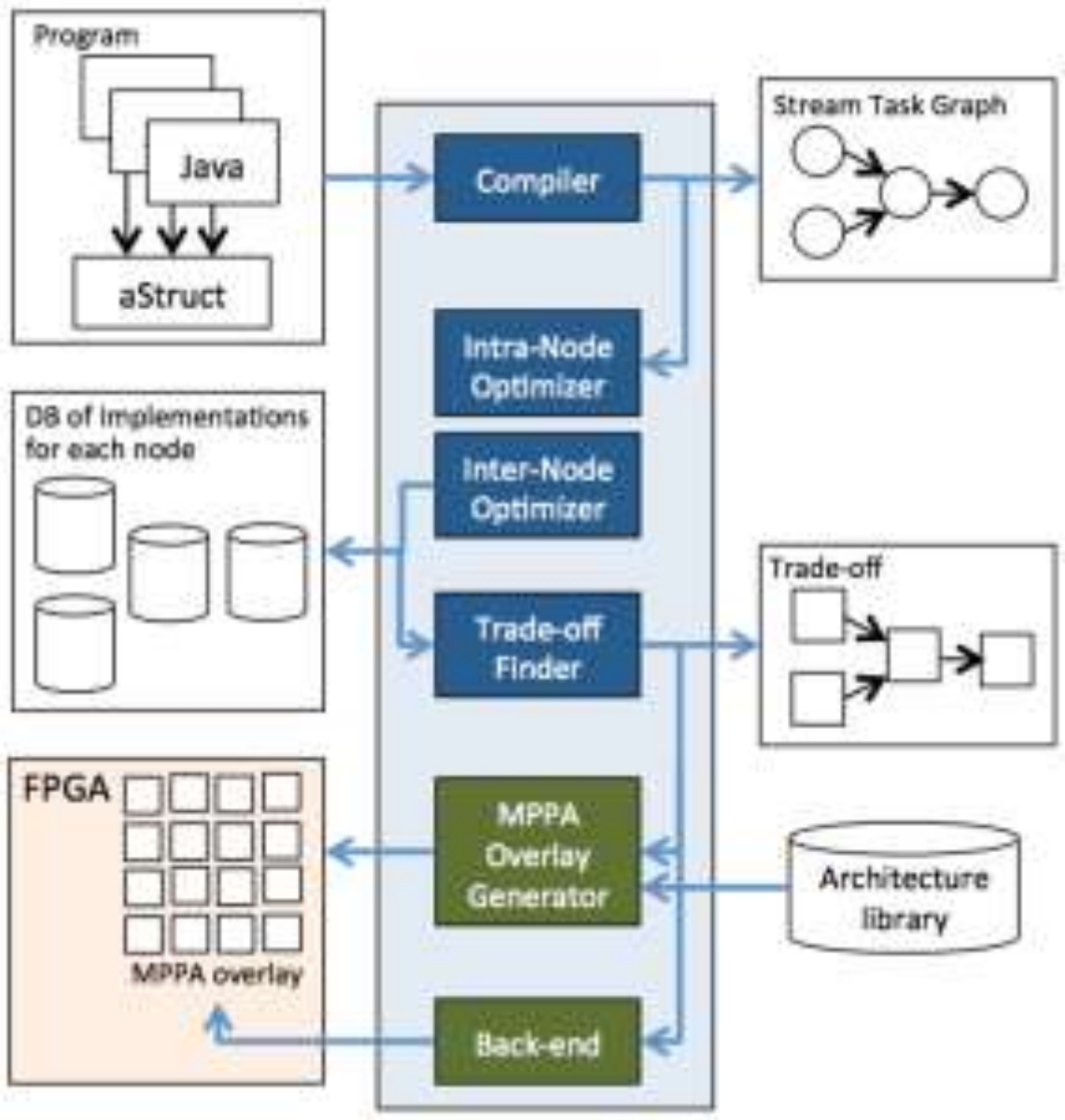

Figure 1: Tool flow

### A. *Finding different implementations*

Consider an application with N composite nodes $f_1, f_2, \dots, f_N$ in its STG. For each composite node $f_m$, our tool tries to find different implementations $P_m^1, P_m^2, \dots, P_m^{S_m}$ where each implementation $P_m^s$ can perform the functionality of $f_m$ with area cost $A(P_m^s)$ and initiation interval $II(P_m^s)$. For node $f_m$ and its implementation $P_m^s$, the minimal input "inverse throughput" $v_{in}(P_m^s)$ and output inverse throughput $v_{out}(P_m^s)$ are calculated as

$$v_{in}(P_m^s) = \frac{II(P_m^s)}{In(f_m)}, v_{out}(P_m^s) = \frac{II(P_m^s)}{Out(f_m)}, \quad (1)$$

Where $In(f_m)$ and $Out(f_m)$ equal the number of data tokens that $f_m$consume on the input data channel and produce on the output data channel during each firing respectively. Note that inverse throughput shows the number of cycles to consume/produce per datum in its input/output channel.

Intra-Node Optimizer and Inter-Node Optimizer modules have been implemented in our tool to automatically find these above-mentioned implementations for each composite node.

#### 1) *Intra-Node Optimizer*

Affine loop transformation strategies in [8,9,10,11,12] have been used in Intra-Node Optimizer to find the maximum degree of parallelism for each composite node. After finding all degrees of parallelism, Intra-Node Optimizer tries to find the best throughput (minimizing the inverse throughput) for each node without considering area cost. Since each operation needs different number of clock cycles to provide its output (different invers throughput), Intra-Node Optimizer (a) expands, (b) combines or (c) pipelines nodes regarding the inverse throughput of operations inside the composite node in order to find an implementation with highest throughput for each composite node.

#### 2) Inter-Node Optimizer

After finding the implementation with the highest throughput for each node, Inter-Node Optimizer starts a clustering operation in order to find different implementations. Each cluster will be mapped to one Processing Element. Inter-Node Optimizer also sends operations back and forth between clusters to find optimum area cost for each overall inverse throughput target.

The example below illustrates how our tool works.

#### 3) Example: N-Body Problem

The traditional N-body Problem simulates a 3D universe, where each celestial object is a body, or particle, with a fixed mass. Over time, the velocity and position of each particle is updated according to interactions with other particles and the environment. In particular, each particle exerts a net force (i.e., gravity) on every other particle. The computational complexity of the basic all-pairs approach we use is O(n2). The run-time of the N-body is dominated by the gravity force calculation, shown below:

$$\overrightarrow{F_{i,j}} = G\frac{M_i M_j}{r^2} = 0.0625\frac{M_i M_j}{\left|\vec{P_i}-\vec{P_j}\right|^3}\left(\vec{P_i}-\vec{P_j}\right), \qquad (2)$$

Where $\overrightarrow{F_{i,j}}$ is the force particle i imposes on particle j, $P_i$ is the position of particle i, and $M_i$ is the size or 'mass' of particle i. When mapping the force calculation, because of the dependencies between instructions in this code, our tool first pipelines it. A simplified 2D pipeline (with latencies) is shown in Figure 2.

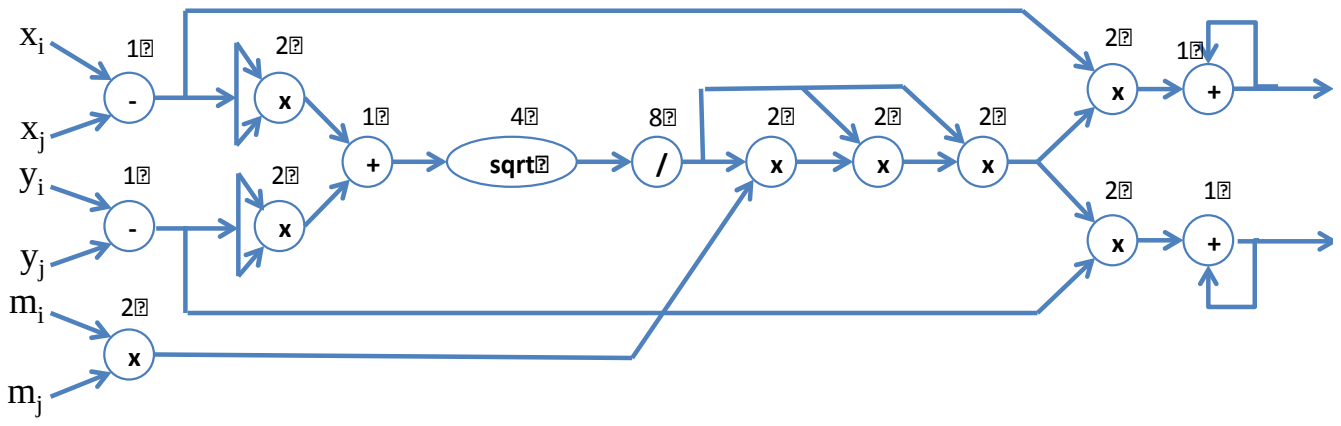

Figure 2: Pipelined Force Calculation

Consider mapping each operation to a simple Processing Element. We get the highest throughput if and only if each operation operates in one clock cycle (inverse throughput=1). As shown in Figure 2, each operation has different inverse throughput. For example, division needs eight cycles to provide its output and, because of dependencies, other operations have to stall for division. It means the best overall inverse throughput we can get for force calculation with this mapping is 8. To get the highest throughput, Intra-Node Optimizer uses an "expansion" approach to parallelize those nodes with high inverse throughput. Figure 3 shows an improved pipeline where inverse throughput equals to 1.

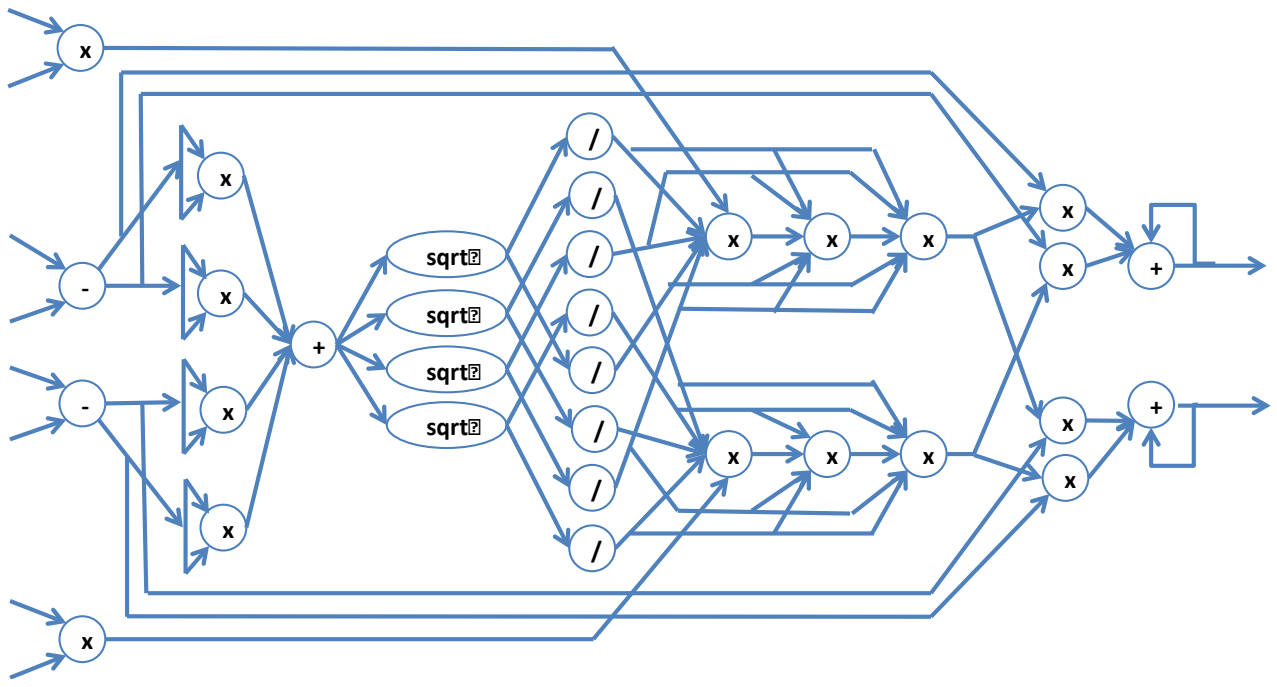

Figure 3: Expanded Force Calculation

After finding the highest possible throughput, Inter-Node Optimizer tries to cluster and combine nodes again to find several implementations with different throughput and area. It means that Inter-Node Optimizer sacrifices the throughput to save area. Inter-Node Optimizer continues this procedure until it assigns the entire composite node $f_m$ to one PE (Area cost = 1). Figure 4 shows inverse throughput and area relation for different implementations of the N-Body function. Here, the inverse throughput varies from 1 to 33. To achieve the best throughput (inverse throughput = 1), we can either replicate the slowest implementation (inverse throughput=33) into 33 copies or use the fastest implementation directly.

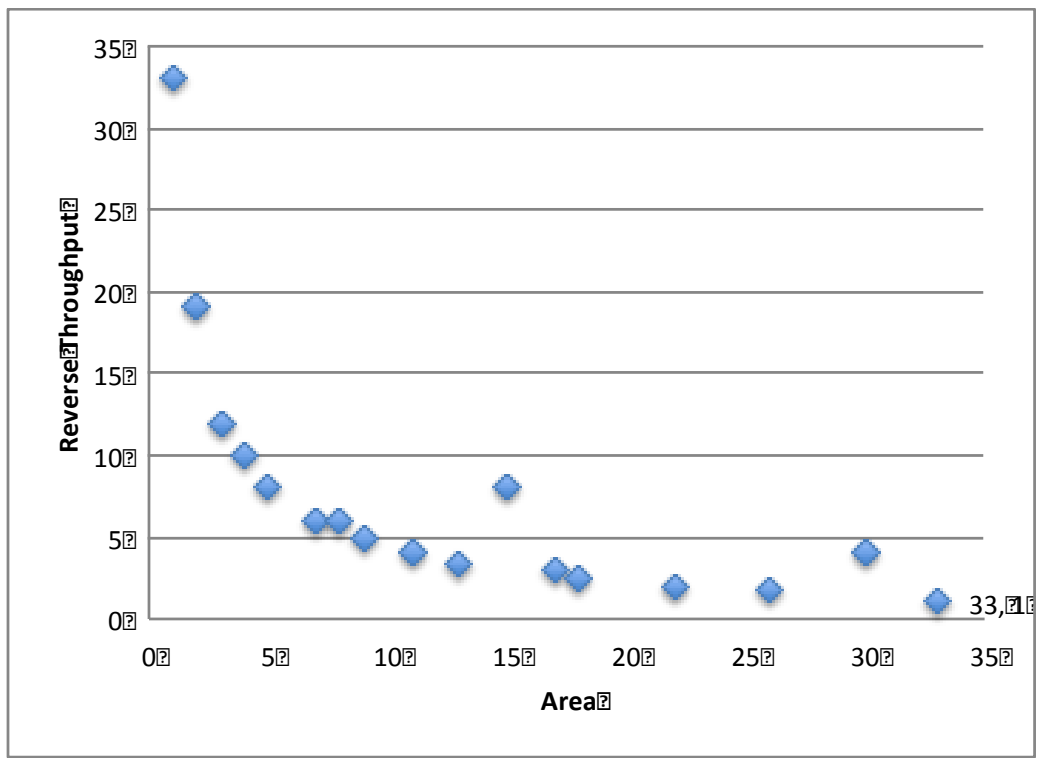

Figure 4: Inverse-Throughput/Area relation for different implementations of N-Body function

### B. Trade-off Finding Formulation and Solutions

Trade-off finding has two different modes in our tool.

- Given an available area on chip $A_C$ and different implementations for each node $f_j$, which implementation $P_j^i$ should be selected and how many replicas $nr_j^i$ are needed in order to minimize application inverse throughput $v_{\mathcal{A}}$ subject to the constraint the application area cost $A_{\mathcal{A}}$ is not bigger than $A_C$.
- Given an inverse throughput target $v_{tgt}$, and different implementations for each node $f_j$, which implementation $P_j^i$ should be selected and how many replicas $nr_j^i$ are needed in order to minimize area cost $A_{\mathcal{A}}$ subject to the constraint the application inverse throughput $v_{\mathcal{A}}$ is not bigger than $v_{tgt}$.

#### 1) Using Integer Linear Programming

The first approach to tackle this problem was defining the problem in terms of Integer Linear Programming (ILP). The goal is to find binary integers $x_{j,1}, x_{j,2}, \ldots, x_{j,S_m}$ indicating the implementations to be selected, and integer $nr_j^i$ indicating number of replicas needed. The formulation of the first problem is:

Minimizing $v_{\mathcal{A}}$ subject to

$$\sum_{j=1}^{N}\sum_{i=1}^{S_m} nr_j^i A(P_j^i) x_{j,i} < A_C \text{ and } \forall j \in \{1,..,N\}: \sum_{i=1}^{S_m} x_{j,i} = 1 \qquad (3)$$

The formulation of the second problem is:

Minimizing $A_{\mathcal{A}}$ subject to

$$\forall j \in \{1,..,N\}: \sum_{i=1}^{S_m} \frac{1}{nr_j^i} v(P_j^i) x_{j,i} < v_j^{tgt} \text{ and } \sum_{i=1}^{S_m} x_{j,i} = 1 \qquad (4)$$

An ILP solver could go through all the possibilities of $x_{j,i}$ and $nr_j^i$ and find the optimum solution for this problem, subject to the constraints. Although ILP solvers can solve these problems, the approach does have two shortcomings:

- Lack of flexibility: the problem must be defined in advance and it's not possible to change the problem's structure while solving it by ILP. In other words, combining or splitting nodes are not possible while using ILP.
- Time inefficient: In our experiments, ILP is usually slower than our heuristic algorithm.

#### 2) Using Heuristic Approach

Before describing our heuristic approach, let us first define Throughput Analysis, Throughput Propagation and Bottleneck Optimizer.

##### a) Throughput Analysis

Each node achieves the maximum output throughput if and only if all its input data are ready when the node expects them. To make the throughput analysis more straightforward we use inverse throughput instead of throughput. To achieve minimum inverse throughput $v_{mo}$, the input data have to be ready with expected inverse throughput $v_{ei}$. We define slack $v_s$ for each channel as

$$v_s = v_{mo} - v_{ei} \qquad (5)$$

Figure 5 shows a simple example in which two nodes A and B nodes are connected together. Node A is a potential bottleneck if it doesn't provide data fast enough to satisfy node B's expectation ($v_{mo} > v_{ei} \Rightarrow v_s > 0$). Node B is a possible bottleneck if node A provides data faster than what node B consumes ($v_{mo} < v_{ei} \Rightarrow v_s < 0$).

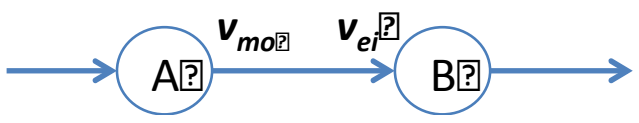

Figure 5: Minimum and expected inverse throughput

Throughput analysis helps us to find bottleneck nodes in a system as well as unnecessary high throughput nodes. Figure 6 shows an example with seven nodes with different $v_{mo}$ and $v_{ei}$. We calculated slack $v_s$ for each channel. As we can see $v_s$ for input channels going to $f_3$ are smaller than $v_s$ for other input channels. Also $v_s$ for output channel from $f_3$ is bigger than $v_s$ for other output channels. This shows $f_3$ is a potential critical bottleneck. To find potential bottlenecks in an application, we define weight $W_m$ for each node $f_m$ as:

$$W_m = \frac{\sum_{j=1}^{N_{out}} v_{sj} - \sum_{i=1}^{N_{in}} v_{si}}{N_{in} + N_{out}} \qquad (6)$$

where $v_{si}$ is the input throughput slacks for incoming channels and $v_{sj}$ is the output throughput slacks for outgoing channels of $f_m$. $N_{in}$ denotes the number of inputs, and $N_{out}$ denotes the number of outputs for node $f_m$. A higher weight means that the node is not able to provide/consume expected outgoing/incoming data to/for its neighbors in most of its channels and the throughput differences between this node and its neighbor are critical.

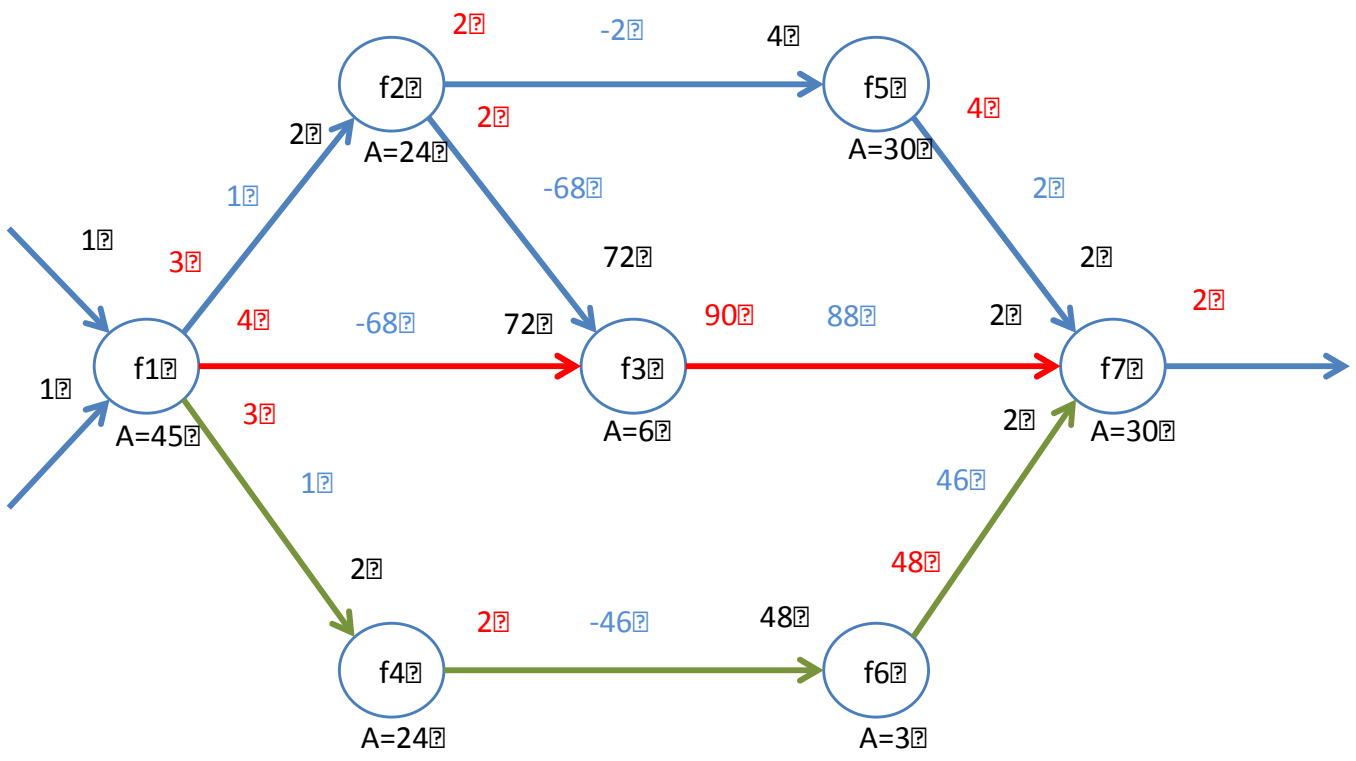

Figure 6: Throughput Analysis

##### b) Application Throughput Propagation and Balancing

Although it seems trade-off finder should select an implementation for each node in order to increase its throughput, increasing throughput of a node will not necessarily increase the overall throughput. For example, as shown in Figure 7, optimizing the block B doesn't increase the throughput of the system (in either case) because the system always has to wait for block A which takes 9 clock cycles. In this case, it might be good to sacrifice the throughput of block B in order to release some area.

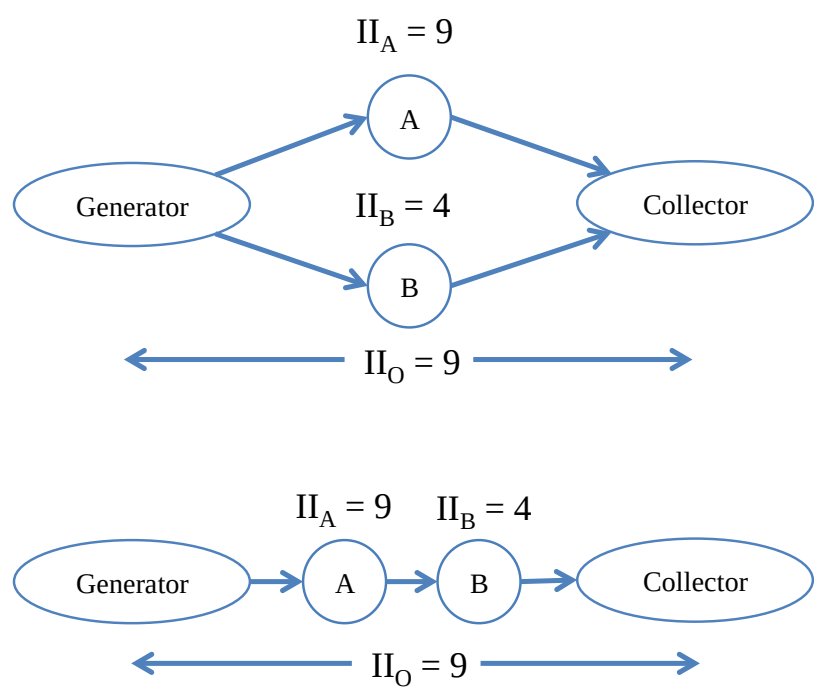

Figure 7: Throughput Propagation and Balancing

To balance the throughput for each node we have to propagate the target inverse throughput to all nodes in the application. For propagating the input inverse throughput target to the output inverse throughput target for a single composite node, we used similar strategy in previous work [7]. For a composite node $f_m$, the number of input and output channels are denoted as $\text{numIn}(f_m)$ and $\text{numOut}(f_m)$ and the inverse throughput target on the input/output channel is denoted as $v^j_{in}(f_m)/v^k_{out}(f_m)$, where $1 \le j \le \text{numIn}(f_m)$ and $1 \le k \le \text{numOut}(f_m)$. Given the input inverse throughput target $v^j_{in}(f_m)$, the output inverse throughput target $v^k_{out}(f_m)$ is calculated as

$$v^k_{out}(f_m) = \frac{\min_j\{v^j_{in}(f_m).In^j(f_m)\}}{Out^k(f_m)} \qquad (7)$$

Throughput propagation helps the Trade-Off Finder to balance the throughput of all nodes in the application.

### *c) Bottleneck Optimizer*

Bottleneck Optimizer is very similar to ILP in sense that it makes replicas of the bottleneck to increase the throughput. However ILP replicates the bottleneck without any attention to its neighbouring nodes that could create opportunities to have a lower area overhead cost. To overcome this not-so-intelligent approach, we propose a method that relies on the fact that each node can send/receive data to/from up to FanIn/FanOut number of nodes without any area overhead cost. If more than FanIn/FanOut numbers of replicas are required, overhead cost is inevitable. In other words, to connect these replicas to successor/predecessor nodes, new fork/join nodes are needed to send/receive data to each replica with a round-robin order are needed.

Let us go through a simple example to show the overall idea in our Bottleneck Optimizer approach. Figure 8.a shows an example which two nodes S with inverse throughput $v_S$ and D with inverse throughput $v_D$ are connected together and node S is sending data to D over a channel. Assume the node D is a bottleneck and we want to match its throughput to node S's throughput. In order to match the throughput we have to make nr replicas of node D which is calculated as

$$nr = \frac{v_D}{v_S} \qquad (8)$$

In order to connect node S to nr replicas of node D we have to use several nodes in between to get data and send it to each replica in round-robin order. Assume each node has FanIn/FanOut equal to nf, which means each node can send data to a maximum of nf nodes. We define $H = \lceil \log_{nf} nr \rceil$, which shows how many layers of nodes need to send data from one node to $nr \le nf^H$ nodes. Assuming $nr = nf^H$ (Figure 8.b), the area overhead $A_O$ for connecting node S to nr replicas of node D is calculated as

$$A_O = \sum_{i=0}^{H-1} nf^i \qquad (9)$$

In our approach, we try to combine nodes together in order to save area overhead. As shown in Figure 8.b in each layer h there are $nf^{h-1}$ nodes with input inverse throughput $v^h_{in}$ and output inverse throughput $v^h_{out}$, which are calculated as

$$v^h_{in} = v_S.nf^{h-1} = \frac{v_D}{nf^{H+1-h}} \qquad (10)$$

$$v^h_{out} = v^h_{in}.nf \qquad (11)$$

So if we can find an implementation S' of node S with inverse throughput equal to $v^h_{in}$, we can combine node S' with nf copies of node D without any area overhead (Figure 8.c) and name it node C with input inverse throughput $v_C$.

$$v_C = \frac{t_D}{nf} \qquad (12)$$

To match the inverse throughput of node C to $v_S$ we have to make $nr'$ replicas of it with area overhead $A'_O$

$$nr' = \frac{nr}{nf} \qquad (13)$$

$$A'_O = \sum_{i=0}^{H-2} nf^i \qquad (14)$$

Assuming that inverse throughput/area relation between node S and node S' is linear, we can save $nf^{H-1}$ nodes. For example in case $nf = 4$, more than 75% overhead area will be saved. As shown in this example we are able to save area by combining nodes, an approach that we cannot model in the ILP formulation.

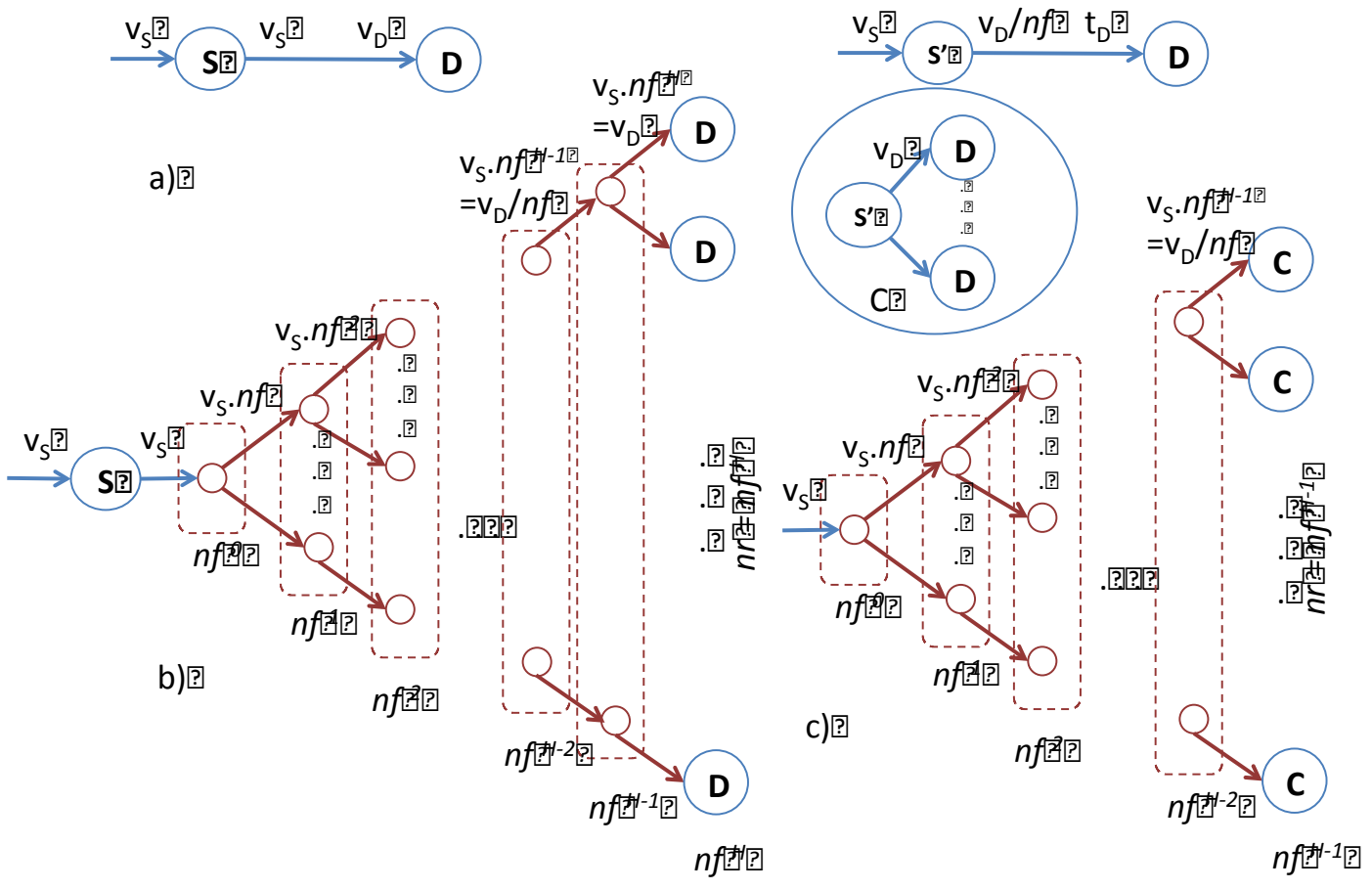

Figure 8: Node combining in Bottleneck Optimizer

Now that the key modules of our heuristic approach have been defined, we will explain how our heuristic approach works below.

### *d) Heuristic Approach Description*

Trade-off Finder starts by selecting an implementation with the highest throughput for each node. Throughput Analyzer analyzes the full application and calculates the expected input inverse throughput and minimum output inverse throughput for each channel. Then, it calculates slacks for all channels and weights for all nodes. Trade-off Finder finds the most critical bottlenecks using the node weights. Next, Trade-off Finder calculates the application area and available area for this implementation. Considering the available area, Trade-off Finder budgets the most critical bottleneck and propagates the throughput to other nodes. After budgeting all nodes, it calculates an approximate area cost for the application considering new throughput for each node. Trade-off Finder accepts an area cost bigger than available area on the chip within a margin. In other words, it overshoots and hopes to release area later from fast nodes in the trading-off finding process. If the approximate area cost is above the margin, Trade-off Finder decreases the target throughput budget and does the same procedure again. After finding a reasonable budgeting, Trade-off Finder starts from the most critical bottleneck on the critical path and uses Bottleneck Optimizer to make replicas of that bottleneck to get a better throughput. Trade-off Finder starts from the optimized bottleneck and goes toward the output until it reaches a node which is located on another critical path. After reaching this node, Trade-off Finder goes backward to visit the other bottleneck and uses Bottleneck Optimizer to match its throughput to satisfy other nodes throughput expectations. The process continues until it balances all the other nodes. Trade-off Finder sees the other nodes in breadth first search order and makes sure that each node doesn't affect nodes in other critical paths. In other words our tool always plays safe, so it selects the implementation to satisfy all channel's throughput for each node.

### *C. MPPA Overlay Finding Problem*

An automated approach for finding area and throughput trade-off for stream applications has been introduced in section 2.1 and 2.2. Finding a good trade-off will not necessarily solve the throughput/area problem. Mapping a steam application to an unsuitable overlay makes the throughput and area trade off finding effort ineffective. Finding a suitable overly is necessary in order to get better throughput while using less area. The current proposed MPPA overlay is an array of processing elements and memories similar to the Ambric architecture. A programmed PE or memory is called an object. Each object is encapsulated and it runs independently at its own speed. All objects intercommunicate through channels. The proposed layout would be a 2D array of tiles. Each tile has Processing Unit (PU) and Memory Unit (MU). PU contains different PEs, which are sharing memory blocks in the MU. MPPA Overlay Generator will decide which type of PE and how many of them should be used in each tile. This problem can be solved by adding another dimension to our trade-off finding problem.

## III. EXPERIMENT RESULTS

Our experiments are carried out in two parts. We first evaluate our strategies of finding different implementations using StreamIt benchmarks [13]. Then, we examine our trade off finding methodology using the JPEG encoder.

### *A. StreamIt*

In order to evaluate our tool, benchmarks in StreamIt benchmark set such as FFT, Filterbank and Autocor are implemented as KPN programming model and fed them to our tool. A simulator has been implemented to validate the results. Our tool was able to find different implementations for each benchmark with different area cost and throughput. The functionality of all the implementations has been verified with the simulator as well. Since all the StreamIT benchmarks are small, we just used them for evaluating the front-end and finding different implementations. We have JPEG, which is more complex in order to examine our trade off finder.

### *B. JPEG*

Figure 10 shows the block diagram of the JPEG compression algorithm. The JPEG compression algorithm contains four major producer/consumer relationships (4 kernels shown in the figure).

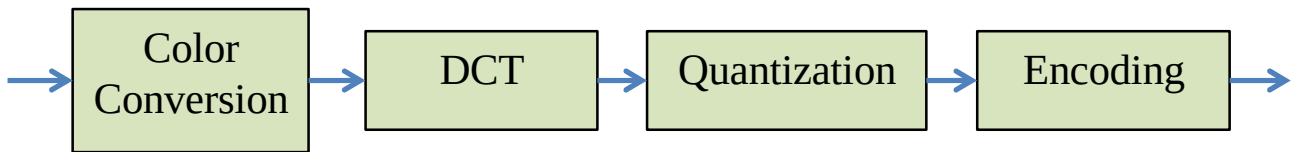

Figure 10: JPEG compression algorithm flow

The tool assumes each kernel is a composite node and, by using Intra-Node Optimizer and Inter-Node Optimizer modules, it finds different implementations for each of them. Our tool found 11 different implementations for Color Conversion and Quantization modules, 17 different implementations for DCT, and only one implementation for Encoding. Table 1 shows a selection of these implementations. Both ILP and Heuristic approaches have been used by our tool in order to find a trade off between area and throughput for different inverse throughput targets for JPEG. Table 2 gives the results generated by these two approaches for given throughput targets. We list the selected implementation and number of replicas for each module. As we can see our heuristic approach finds better area/throughput trade-off compare of the ILP approach. For example, for an inverse throughput target of 2, our heuristic approach used 37% less area compare to ILP. The ILP solver we use is GLPK [14] and the area cost unit is primitive node which can be implemented as a CLB in FPGA.

TABLE I. IMPLEMENTATION LIBRARY FOR JPEG ENCODER

| Module | Color Conversion | | | | DCT | | | | | Quantization | | | | | Encoding |
|---|---|---|---|---|---|---|---|---|---|---|---|---|---|---|---|
| Version | v1 | v2 | v3 | v4 | v1 | v2 | v3 | v4 | v5 | v1 | v2 | v3 | v4 | v5 | v1 |
| v | 1 | 2 | 4 | 8 | 1 | 2 | 4 | 6 | 32 | 1 | 2 | 4 | 8 | 128 | 512 |
| Area | 512 | 256 | 128 | 64 | 800 | 400 | 224 | 160 | 50 | 512 | 256 | 128 | 64 | 4 | 22 |

TABLE 2. ILP VS. OUR HEURISTIC METHOD

| Method | v | Color Conversion | | DCT | | Quantization | | Encoding | | Fork/join Overhead | Total Area |
|---|---|---|---|---|---|---|---|---|---|---|---|
| | | impl | rep | impl | rep | impl | rep | impl | rep | | |
| ILP | 1 | v1 | 1 | v1 | 1 | v1 | 1 | v1 | 512 | 10880 | 23968 |
| Heuristic | | v1 | 1 | v5 | 32 | v5 | 128 | v1 | 512 | 640 | 13888 |
| ILP | 2 | v2 | 1 | v2 | 1 | v2 | 1 | v1 | 128 | 5376 | 11920 |
| Heuristic | | v2 | 1 | v5 | 16 | v5 | 64 | v1 | 128 | 256 | 7456 |
| ILP | 4 | v3 | 1 | v3 | 1 | v3 | 1 | v1 | 64 | 2688 | 5984 |
| Heuristic | | v3 | 1 | v5 | 8 | v5 | 32 | v1 | 64 | 128 | 3600 |
| ILP | 8 | v4 | 1 | v4 | 1 | v4 | 1 | v1 | 32 | 1280 | 2976 |
| Heuristic | | v4 | 1 | v5 | 4 | v5 | 16 | v1 | 32 | 0 | 1736 |

## IV. FUTURE WORK

We mentioned in Section 2 that we will investigate finding and generating suitable MPPA overlays for the found trade-off. Different types of PE as a spectrum from a basic PE to a very complex PE will be implemented as a library. The above-mentioned MPPA Overlay Generator will select suitable PEs for the application's needs. This selecting process adds another dimension to our trade-off problem (area, throughput and target architecture), which can be tackled with both ILP and heuristic approaches.

We mentioned in Section 2.3 that in replication processing, using fork and join modules is inevitable for a large number of replicas. Since most of nodes in fork and join modules are just passing data, they are underutilized most of the time. We may be able to use resource sharing to use the stalling time of underutilized nodes for running other independent computation.

## V. CONCLUSION

In this paper, we studied the HLS problem of automatically finding area/throughput trade-off of streaming applications being mapped onto MPPA overlays. We introduce a high-level synthesis tool that compiles an stream application written in Java as a streaming task graph, partitions it into composite nodes, finds all degrees of parallelism for each, uses different approaches in order to find different implementations for each node, and finally finds a good trade off between area and throughput. Our approach differs from existing approaches because 1) it automatically investigates partitioning and finding different implementations, and 2) it combines module selection and replication methods with node combining and splitting in order to automatically find a better area/throughput trade-off. This approach has been verified with small designs in StreamIt and a few larger designs like the JPEG encoder. This study is also our starting point for finding and generating suitable MPPA overlays for stream applications.